\documentclass[twocolumn]{aastex62}

\usepackage{amsmath,amssymb}
\usepackage{epstopdf}
\usepackage{ulem}
\usepackage{multirow}
\usepackage{soul}
\usepackage{tabularx}

\shorttitle{DSA at oblique shocks}
\shortauthors{van Marle et al.}

\newcommand{\ajvm}{\textcolor{black}}
\newcommand{\ab}{\textcolor{black}}
\newcommand{\mponn}{\textcolor{black}}

\newcommand{\secrev}{\textcolor{black}}

\def\ee{\end{equation}}
\def\be{\begin{equation}}
\def\l{\left}
\def\r{\right}

\newcommand{\omci}{\Omega_\mathrm{i}}

\newcommand{\ms}{M_\mathrm{s}}

\newcommand{\ma}{M_\mathrm{A}}
\newcommand{\mi}{m_\mathrm{i}}
\newcommand{\me}{m_\mathrm{e}}
\newcommand{\lse}{\lambda_\mathrm{se}}
\newcommand{\lsi}{\lambda_\mathrm{si}}
\newcommand{\vsh}{v_\mathrm{sh}}

\begin{document}

\title{Diffusive shock acceleration at oblique high Mach number shocks.}

\author[0000-0002-2387-4515]{Allard Jan van Marle}
\affil{Laboratoire Univers et Particules de Montpellier (LUPM) Universit\'e Montpellier, CNRS/IN2P3, CC72, Place Eug\`ene Bataillon, F-34095 Montpellier Cedex 5, France}

\correspondingauthor{Artem Bohdan}
\email{artem.bohdan@desy.de}

\author[0000-0002-5680-0766]{Artem Bohdan}
\affil{Deutsches Elektronen-Synchrotron DESY, Platanenallee 6, 15738 Zeuthen, Germany}

\author{Paul J. Morris}
\affil{Deutsches Elektronen-Synchrotron DESY, Platanenallee 6, 15738 Zeuthen, Germany}

\author[0000-0001-7861-1707]{Martin Pohl}
\affil{Deutsches Elektronen-Synchrotron DESY, Platanenallee 6, 15738 Zeuthen, Germany}
\affil{Institute of Physics and Astronomy, University of Potsdam, 14476 Potsdam, Germany}

\author{Alexandre Marcowith}
\affil{Laboratoire Univers et Particules de Montpellier (LUPM) Universit\'e Montpellier, CNRS/IN2P3, CC72, Place Eug\`ene Bataillon, F-34095 Montpellier Cedex 5, France}

\begin{abstract}
The current paradigm of cosmic ray (CR) origin states that the most part of galactic CRs is produced by supernova remnants.  The interaction of supernova ejecta with the interstellar medium after supernova's explosions results in shocks responsible for CR acceleration via diffusive shock acceleration (DSA). We use particle-in-cell (PIC) simulations and a combined PIC-magnetohydrodynamic (PIC-MHD) technique to investigate whether DSA can occur in oblique high Mach number shocks. Using the PIC method, we follow the formation of the shock and determine the fraction of the particles that gets involved in DSA. Then, with this result, we use PIC-MHD simulations to model the large-scale structure of the plasma and the magnetic field surrounding the shock and find out whether or not the reflected particles can generate the upstream turbulence and trigger DSA. We find that the feasibility of this process in oblique shocks depends strongly on the Alfvénic Mach number, and the DSA process is more likely triggered at high Mach number shocks.

\end{abstract}


\section{Introduction}\label{introduction}
Collisionless shocks are commonly observed when plasmas move with super-sonic velocities, for example in planetary systems, supernova remnants (SNRs) and jets of active galactic nuclei. In such flows, collective particle interactions form a shock layer on kinetic plasma scales much smaller than the collisional mean free path. In the shock transition a part of the bulk kinetic energy is converted into energies of thermal particles and electromagnetic fields through wave-particle interactions. The microphysics of these processes is still not fully understood, even if it has been recently the object of several numerical studies, especially in the high Mach number regime \citep[e.g.][]{Marcowith:2020,2020PrPNP.11103751P}. This regime will be partly covered in this study.

Diffusive shock acceleration (DSA), a process entering the category of first-order Fermi acceleration, is the process by which astrophysical shocks accelerate charged particles to relativistic speeds. DSA requires the magnetic field near the shock front to reflect the particle, leading to repeated shock-crossings with the particle gaining energy at each crossing \citep[e.g.][]{Bell:1978,Blandford:1978,Drury:1983}.
DSA is a self-sustaining process because the presence of high-energy particles triggers instabilities in the magnetic field, which in turn allow the magnetic field to reflect the particles more efficiently \citep[e.g.][]{Bell:1978,Bell:2004}. 
DSA has been explored numerically using the particle-in-cell (PIC) method, as well as the PIC-hybrid method, whereby the ions are treated as particles and the electrons as a fluid. Simulations of this type seem to indicate that although this process is effective in the case of (quasi-)parallel shocks where the magnetic field is aligned with the direction of  motion, it becomes ineffective once the angle between the magnetic field and the shock exceeds approximately 50$^{\rm o}$ \citep{Caprioli:2014a, Caprioli:2014b, Caprioli:2014c} \footnote{Notice that most of works on DSA including the present one assume shocks are propagating into an homogeneous medium, effects of a non-monotonous shock are discussed in \citet{Hanusch2019}. \ab{Most simulations also assume that the pre-shock medium is fully ionized. Simulations for a partially ionized medium \citep{2013PhRvL.111x5002O,2016ApJ...827...36O} show an increased injection rate, allowing for DSA even in oblique shocks.} Also studies by \cite{Kumar2021} demonstrate that  DSA can be triggered by the shock emitted waves at highly oblique shocks.}
However, different results where obtained by \cite{vanMarle:2018}, using a combined PIC-MHD approach. This method, which treats the thermal gas as a fluid, but non-thermal ions as individual particles, showed that, given a sufficient injection rate, even for magnetic obliquity up to 70$^o$ shocks can accelerate particles through the DSA process. Several factors were believed to contribute to this discrepancy between PIC-MHD and PIC or PIC-hybrid methods:  
\begin{enumerate}
\item Time-period covered by the simulation. The DSA occurring in PIC-MHD simulations is only effective after more than 200 ion-cyclotron times ($\omega_{\rm ci}^{-1}$) and it takes more than 500 ion-cyclotron times to fully develop whereas published hybrid simulations where limited to scales of the order of 200 ion-cyclotron times. Notice that at times $\sim 200~\omega_{\rm ci}^{-1}$ both PIC-hybrid and PIC-MHD simulations find similar results, namely that particles are accelerated by the shock drift acceleration (SDA).
\item Effective number of particles per cell. DSA requires an upstream current of CRs in order to develop. Because only a small fraction of the particles that cross the shock is reflected, a large particle population is required to produce sufficient non-thermal particles in the upstream medium. 
\item Simulation box size. \cite{vanMarle:2018} showed that the relevant upstream instabilities operate on long wavelengths, larger than the box-size of hybrid simulations. 
\end{enumerate}

In order to address these discrepancies, \citet{Haggerty:2019} present an improved version of the PIC-Hybrid code including relativistic ions dynamics. Long simulations, with larger box sizes and equivalent non-thermal particle statistics as in PIC-MHD simulations were performed for Mach number 30 shocks with a magnetic obliquity of $70^o$. The authors confirm that DSA does not develop at this obliquity angle. This leaves a final possibility: the injection rate of non-thermal particles. Because the PIC-MHD method treats the thermal plasma as a fluid, it cannot model the process by which particles start to deviate from thermal equilibrium as they pass through the shock. Instead, it relies on a ad-hoc description by which a fixed fraction of the gas crossing the shock is converted into non-thermal particles. \citet{vanMarle:2018} assumes that the injection rate for a high obliquity shock was equal to that of a quasi-parallel shock. However, as studies shown in \citet{Caprioli:2014a} and this paper shows, this is not the case. The injection rate for high obliquity shocks is substantially lower than for a quasi-parallel shock with a similar Mach number. The question is whether this lower fraction is sufficient to trigger the kind of upstream instability that is required to initiate DSA. Another aim is to test, using a full PIC approach, the injection rate obtained with the PIC-Hybrid method as its level has a strong impact on the development of DSA and streaming instabilities. An analytical approximation of the relevant timescales in Appendix~A of \citet{Vanmarle:2020} shows that the ability of particles to distort the upstream magnetic field depends, among other factors, on the local particle density. If it is too low, the thermal gas and the magnetic field can resist the force that the particle are exerting on the field. Notice that this is not the same as the criterion for triggering the streaming instability. It is possible to distort the magnetic field without causing the thermal gas to become unstable. Instead, the magnetic field is distorted by the presence of non-thermal particles through the Lorentz force, which reduces the particle energy. Subsequent waves of particles are then accelerated by the distortions at the expense of reducing the distortion of the magnetic field. This creates a cyclic pattern by which one wave of particles distorts the field while losing energy and the next smooths the field while accelerating. A similar pattern was observed for high-$\beta$ shocks at sonic Mach numbers below 2.85 \citep{Vanmarle:2020}.

In this paper, we further explore the parameter space of high-obliquity shocks to investigate which, if any, semi-perpendicular shocks can generate a sufficient number of non-thermal particles to trigger upstream distortions of the magnetic field and to eventually initiate DSA. We do so by performing 2D PIC simulations to determine the injection rate of supra-thermal particles at the shock as a function of shock velocity, Mach number, and obliquity. Once the injection rate is properly defined, we perform corresponding 2D PIC-MHD simulations to follow the large-scale, long-term evolution of the plasma and the magnetic field near the shock. The combination of the two techniques is used to cover a wider range of plasma beta and its impact on the shock acceleration efficiency. Then, PIC-MHD simulations are used to investigate high Alfv\'enic Mach number shocks, a regime barely addressed in hybrid simulations.

We present a short description of simulation setup in Section~\ref{sec:setup}. The results are presented in Section~\ref{sec:results}. The discussion and conclusions are given in Sections~\ref{sec:discussion} and ~\ref{sec:conclusions}, correspondingly.

\section{Numerical methods and setups} \label{sec:setup}
\subsection{PIC simulations}\label{sec:pic}

\begin{figure*}[htp]
\centering
\includegraphics[width=0.49\linewidth]{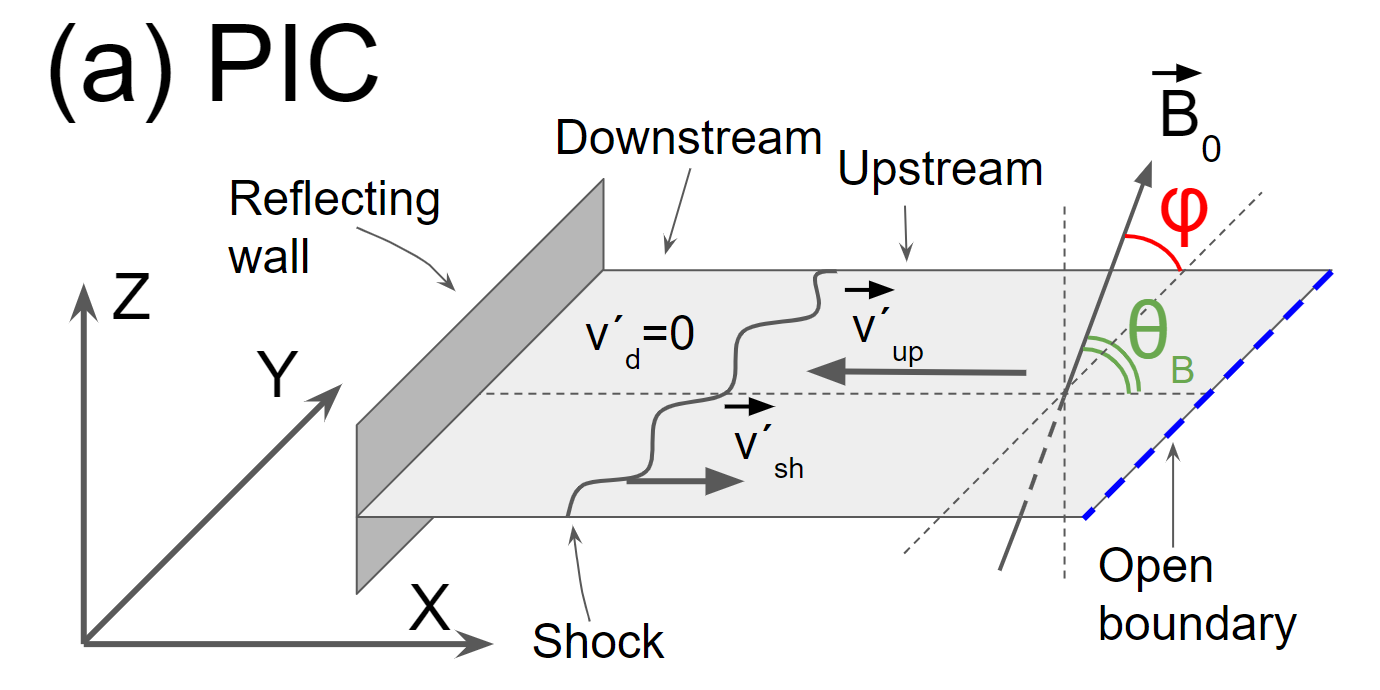}
\includegraphics[width=0.49\linewidth]{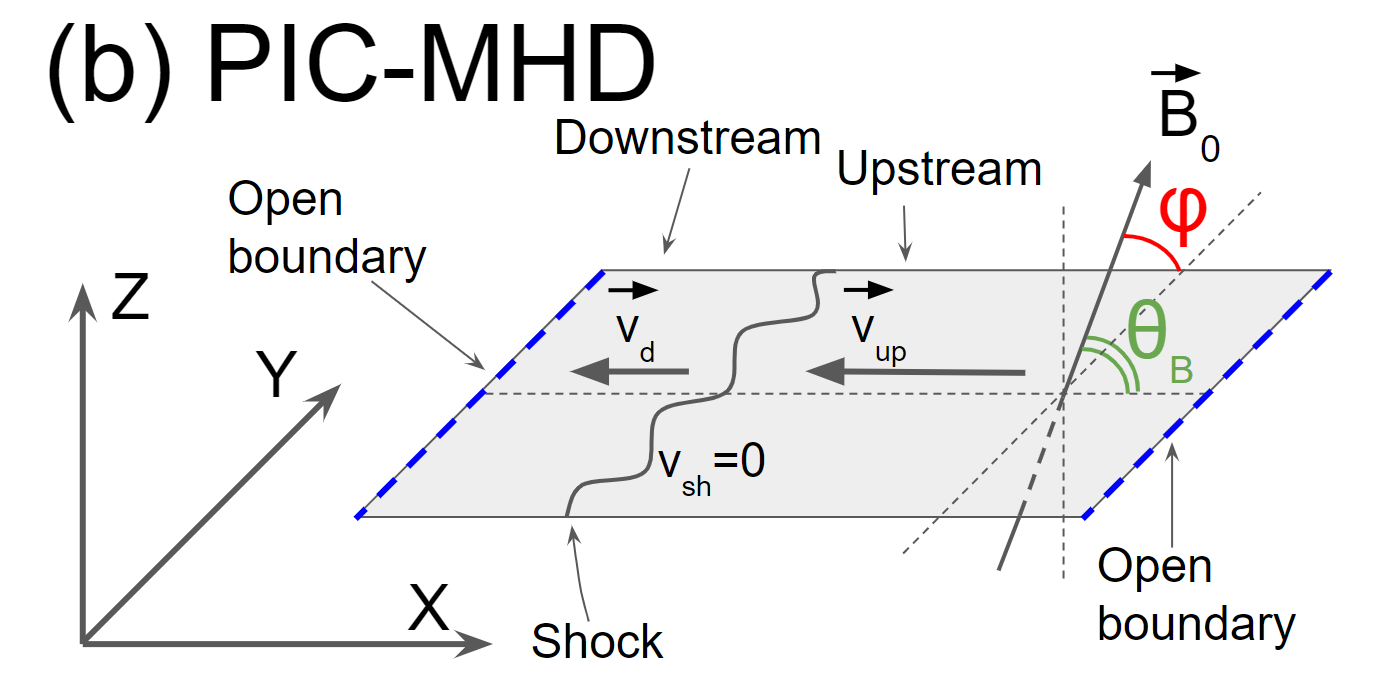}
\caption{Sketch of the setups for PIC (a) and PIC-MHD (b) shock simulations.}
\label{setup_shock_pic}
\end{figure*}

PIC simulations are performed using the modified version of the relativistic electromagnetic TRISTAN code \citep{Buneman1993} with MPI-openMP hybrid parallelization \citep{2008ApJ...684.1174N} and the particle sorting optimization \citep{10.1007/978-3-319-78024-5_15}. The Vay solver \citep{Vay2008} is used to update particle positions. 
The triangular-shape-cloud particle shapes (the second-order approximation) and \cite{Friedman1990} filter for electric and magnetic fields are used to suppress the numerical grid-Cherenkov short-wave radiation. The numerical model used in our simulations has been extensively tested, and all individual upstream energy components are conserved to better than 5\% accuracy over the entire time of simulation.

We initialize shocks using the reflecting-wall setup, where the left boundary is reflective, the right boundary is open, and periodic boundary conditions are applied in y-direction. The electron-ion upstream plasma moves in negative $x$-direction with the absolute velocity $v_0$ (Fig.~\ref{setup_shock_pic}(a)). It collides with the conducting wall, and further interaction of the reflected and the upstream plasma flows provides a shock propagating in positive $x$-direction. To prevent an initial transient caused by this artificial contact discontinuity between the reflecting wall and the plasma slab, we introduce a drift current to ions close to the reflecting wall in the same way it is discussed in \citet{2016ApJ...820...62W}.

       \begin{centering}
\begin{table*}[!t]
      \caption{PIC Simulation Parameters}
         \label{table-param_pic}
\begin{tabular}{p{0.08\linewidth}p{0.05\linewidth}p{0.14\linewidth}p{0.06\linewidth}p{0.06\linewidth}p{0.06\linewidth}p{0.06\linewidth}p{0.1\linewidth}p{0.1\linewidth}}
\hline
\hline
\noalign{\smallskip}
Name  & $\theta_B$ & in/out-of plane & $\ma$ & $V_\mathrm{sh}/c$ & $\beta_p$ & $\omega_{\rm pe}/\Omega_{\rm e}$ & $N_{\rm inj}/N_{\rm 0}$ & $U_{\rm inj}/U_{\rm sh}$ \\
\noalign{\smallskip}
\hline
\noalign{\smallskip}
PIC01   &   65 & in-plane & 20 & 0.133 & 1 & 21.2  & 0 &  0  \\ 
PIC02   &   65 & out-of-plane & 20 & 0.133 & 1 & 21.2  &  $2.8 \times 10^{-6} $ &$ 5.8 \times 10^{-5}$   \\ 
PIC03   &  60 & in-plane & 20 & 0.264 & 1 & 10.6  &  $2.8 \times 10^{-6} $&$ 6.4 \times 10^{-5}  $ \\ 
PIC04    &   60 & out-of-plane & 20 & 0.264 & 1 & 10.6  & $ 1.3 \times 10^{-5}$ &$ 2.5 \times 10^{-4} $  \\
PIC05    &   60 & out-of-plane & 30 & 0.264 & 1 & 15.9  &  $4.3 \times 10^{-5} $&$ 5.5 \times 10^{-4}$   \\
PIC06    &   60 & out-of-plane & 30 & 0.133 & 1 & 31.9  &  $9 \times 10^{-5} $&$  1.1 \times 10^{-3}$  \\
PIC07    &  60 & in-plane & 20 & 0.133 & 1 & 21.2  &  $8.4 \times 10^{-6} $&$ 1.3 \times 10^{-4} $ \\
PIC08    &  60 & out-of-plane & 20 & 0.133 & 1 & 21.2  & $ 6.5 \times 10^{-5}  $&$  8.9 \times 10^{-4}  $\\
PIC09    &  60 & out-of-plane & 20 & 0.133 & 10 & 21.2  & $ 3.7 \times 10^{-5}  $&$  5.7 \times 10^{-4}  $\\
PIC10    &  60 & out-of-plane & 20 & 0.133 & 0.1 & 21.2  & $ 3.3 \times 10^{-5}  $&$ 4.5 \times 10^{-4}$  \\
PIC11    &  55 & in-plane & 20 & 0.133 & 1 & 21.2  &  $4 \times 10^{-4} $&$  4.2 \times 10^{-3} $ \\
PIC12    &  55 & out-of-plane & 20 & 0.133 & 1 & 21.2  & $ 4.1 \times 10^{-4} $&$  4 \times 10^{-3} $ \\
PIC13    &  50 & in-plane & 20 & 0.133 & 1 & 21.2  &  $2.7 \times 10^{-3}$ &  0.02  \\
PIC14    &  50 & out-of-plane & 20 & 0.133 & 1 & 21.2  & $ 2.6 \times 10^{-3} $ &  0.018  \\
PIC15    &  45 & in-plane & 10 & 0.133 & 1 & 10.6  &  0.016 &  0.086  \\
PIC16    &  45 & in-plane & 20 & 0.133 & 1 & 21.2  &  0.013 &  0.068  \\
PIC17    &  45 & out-of-plane & 20 & 0.133 & 1 & 21.2  &  0.022 &  0.12  \\
\noalign{\smallskip}
\hline
\end{tabular}
\tablecomments{Setup of the PIC simulations. Listed are: 
the shock obliquity, in-plane or out-of-plane configuration, the shock speed, the Alfv\'enic Mach number, $\ma$, and the plasma parameter $\beta_{\rm p}$. \ab{The ion-to-electron mass ratio is equal to 50 for all PIC simulations.}
The ion injection density and energy density fractions are reported in the last two columns (see section \ref{sec:picres_sub2} for details).} 
\end{table*}
       \end{centering}

The large-scale upstream magnetic field, $B_0$, makes an angle $\theta_{B}$ with the shock normal (or positive $x$-direction) and an angle $\varphi$ with the simulation plane. Simulations where $\varphi=0^o$ and $\varphi=90^o$ are referred to as \emph{in-plane} and \emph{out-of-plane}, correspondingly. Shock simulations are performed in 2D3V configuration which follows two spatial coordinates and all three components of the particle velocities and electromagnetic fields. Therefore the adiabatic index is $\Gamma=5/3$ and the expected compression ratio is about 4. The resulting shock speed either equals $\vsh'=0.067c$ or $0.033c$ in the simulation (\emph{downstream}) reference frame and $\vsh=0.263c$ or $0.133c$ in the \emph{upstream} reference frame. 

The Alfv\'en velocity is defined as $v_{\rm A}=B_{\rm 0}/\sqrt{\mu_{\rm 0} \rho}$ with $\rho =N_e\me+N_i\mi$, where $\mu_{\rm 0}$ is the vacuum permeability, \ab{$\mi$ and $\me$ are the ion and the electron masses}, $N_i$ and $N_e$ are the ion and the electron number densities, which are 40 particles per cell for each species. The sound speed reads $c_{\rm s}=(\Gamma k_BT_{\rm i}/\mi)^{1/2}$, where $k_B$ is the Boltzmann constant and $T_{\rm i}$ is the ion temperature that initially is equal to the electron temperature defined as $k_BT_i=m_i v_{th}^2/2$, where $v_{th}$ is defined as the most probable speed of the upstream plasma particles in the upstream reference frame. The plasma beta is $\beta_{\rm p} = 2\mu_{\rm 0} N_{\rm i} k_{\rm B} T_{\rm p}/ B^2$. The Alfv\'enic, $\ma=\vsh/v_{\rm A}$, and sonic, $\ms=\vsh/c_{\rm s} = \sqrt{2/\Gamma\beta_{\rm p}} \ma$, Mach numbers of the shocks are defined in the conventional \emph{upstream} reference frame. We list in Table~\ref{table-param_pic} the parameters $\beta_{\rm p}$ and $\ma$, then $\ms$ can easily be deduced.

The ratio of the electron plasma frequency, $\omega_{\rm pe}=\sqrt{e^2N_e/\epsilon_0\me}$, to the electron gyrofrequency, $\Omega_{\rm e}=eB_0/\me$, is in the range of $\omega_{\rm pe}/\Omega_{\rm e}=10.6-31.9$. Here, $e$ is the electron charge, and $\epsilon_0$ is the vacuum permittivity.
The electron skin depth in the upstream plasma is constant for all runs and equals $\lse=8\Delta$, where $\Delta$ is the size of a grid cell. The ion skin depth is defined as $\lsi=\sqrt{\mi/\me}\lse$ \ab{and $\mi/\me = 50$ for all PIC simulations}. The spatial scale is given in terms of the ion upstream gyroradius, $r_{\rm gi} = m_{\rm i} v_{\rm sh}/(eB_0)$. The simulation time-step is $\Delta t=1/16\,\omega_{\rm pe}^{-1}$. The time scales are given in terms of the upstream ion Larmor frequency, $\Omega_{\rm ci}=eB_0/\mi$. The simulation time is $T_\mathrm{sim} \approx 30 \Omega_i^{-1}$ for all simulations, except run PIC15 where $T_\mathrm{sim} \approx 68 \Omega_i^{-1}$. The simulation time of PIC simulations is limited by the computational expense, especially at high Alfv\'enic Mach numbers and low shock velocities. Also, if a simulation is too long, artificial heating can substantially change the upstream conditions.

\subsection{PIC-MHD simulation setup}\label{sec:picmhd}

       \begin{centering}
  \begin{table*}[!t]
      \caption{PIC-MHD Simulation Parameters}
         \label{table-MHDparam}
\begin{tabular}{p{0.14\linewidth}p{0.06\linewidth}p{0.06\linewidth}p{0.06\linewidth}p{0.14\linewidth}p{0.14\linewidth}p{0.1\linewidth}}
 \hline
 \hline
  Name         &    $\ma$  & $\theta_B$[deg] & $V_\mathrm{sh}/c$  & $N_{\rm inj}/N_0$ & special physics & Notes \\
  \hline
  PICMHD01A     &    20  &  60           & 0.05     &  $5.0\times10^{-5}$ & \  & No~DSA \\
  PICMHD01B     &    20  &  60           & 0.1      &  $2.5\times10^{-5}$ & \   & No~DSA \\
  PICMHD02     &     20  &  60           & 0.1      &  $5\times10^{-5}$ &  \  & No~DSA\\
  PICMHD03     &     20  &  60           & 0.1      &  $1\times10^{-4}$ & \   &  DSA\\
   \hline
   PICMHD04    &     30 &  60            & 0.05      &  $5.0\times10^{-5}$ &  \  & start of DSA\\
  PICMHD05A     &     50  &  60          & 0.05      &  $5.0\times10^{-5}$ &  \  & DSA \\
  PICMHD05B     &     50  &  60          & 0.05      &  $5.0\times10^{-5}$ &  out-of-plane & No DSA \\
  PICMHD06     &     100  &  60           & 0.05     &  $5.0\times10^{-5}$ & \  &DSA\\
  PICMHD07     &     300  &  60           & 0.05     &  $5.0\times10^{-5}$ & \   & DSA\\
  \hline  
  PICMHD08     &      50 &   60          & 0.05      & $5.0\times10^{-5}$ &   $\beta$=0.03  &  DSA\\
  PICMHD09     &      50 &   60          & 0.05      & $5.0\times10^{-5}$ &   $\beta$= 0.1   & DSA\\
  PICMHD10     &      50 &   60          & 0.05      & $5.0\times10^{-5}$ &  $\beta$= 10    &  DSA \\
  PICMHD11    &       50 &   60          & 0.05      & $5.0\times10^{-5}$ &   $\beta$= 30    &  DSA \\ 
  \hline
  PICMHD12    &       50 &   55          & 0.05      & $4.0\times10^{-4}$ &   $\beta$=  30    &  DSA \\
  \hline
\end{tabular}
 
\tablecomments{List of setups of PIC-MHD simulations. The columns show the Alfv\'enic Mach number, $\ma$, the magnetic field obliquity, $\theta_{\rm B}$, the shock speed in units of the speed of light, and the cosmic-ray injection fraction. All simulations have an in-plane magnetic field and $\beta=1$, unless specified otherwise. \ab{N.B. In PIC-MHD, the particles are injected isotropically in the post-shock rest-frame. Therefore we inject $2N_{\rm inj}/N_0$ to get the intended number of reflected particles.}} 
   \end{table*}
  \end{centering}

The PIC-MHD setup is different from the PIC model. In this model the simulations are run in the frame of reference of the shock rather than the frame of reference of the downstream medium, using a similar approach as used in \citet{vanMarle:2018} and \citet{Vanmarle:2020}. The setup is initialized by starting from the analytical solution for a standing shock as derived through the Rankine-Hugoniot conditions for a magnetized thermal gas. 
At the start of the simulation, the grid is filled with a thermal gas according to the specific shock conditions of that simulation. Once the simulation starts, non-thermal particles are introduced at the shock front, according to the injection rates derived from the corresponding PIC simulations. The particles are given an absolute velocity of twice the pre-shock velocity of the thermal gas and an isotropic velocity distribution in the post-shock medium. \ab{Because we use an isotropic velocity distribution in the post-shock rest-frame, we inject twice the number of non-thermal particles. This gives us the desired \emph{reflected} particle fraction}. \ab{As will be shown by the PIC results (See Sect.~\ref{sec:picres}), there is a difference in injection rates between simulations with the magnetic field in the plane of the simulation, and those where the magnetic field is out of the plane. This indicates that the 2D results are not fully representative of the 3D reality. Therefore, we choose an injection rate in between the two results two reflect the 3D nature of the problem.}

We use a 2D simulation box\footnote{\citet{vanMarle:2019} performed a 3D study of particle acceleration at shocks.} that covers $240\times30\,r_g$, with $r_g$ the gyro radius defined by the particle injection velocity and the upstream magnetic field. The grid has a base resolution of $240\times30$ grid cells with the nested OCTREE adaptive mesh allowing for 3 additional levels of refinement which gives us an effective grid of $1920\times240$ cells.\footnote{Note that an insufficient length of the simulation box in the direction of flow may change the outcome of the simulation. See Appendix~B for further explanation.}
The thermal gas flows in from the outer boundary of the x-axis, passes through the shock halfway along the x-axis, and leaves through the inner x-boundary. Both x-boundaries are transparent to particles, allowing the particles to leave the simulation when they reach them. The y-boundaries are periodic for both thermal gas and particles. The magnetic field in the thermal gas makes an angle $\theta_{\rm B}$ with the flow in the plane of the simulation. \ajvm{In order to investigate the occurrence of DSA, we need to be able to follow the particles as they move along the field lines. Therefore, the PIC-MHD simulations have to be performed with the magnetic field in the plane of the simulation. We have performed one PIC-MHD simulation with the field out-of-plane for comparison to the PIC simulations.} The simulation parameters for the different set-up are summarized in Table~\ref{table-MHDparam}. 

Common runs between PIC and PIC-MHD techniques are: PIC-MHD01B and PIC07. 
The PIC-MHD code does not handle relativistic MHD, hence it can not handle flows with speeds much in excess of 0.1c;  this explains why most of the simulations are done for $V_{\rm sh}=0.05$c. However, as we discuss below the injection rates are not strongly dependent on the shock speed at non-relativistic speeds. A strong dependence does develop at relativistic speeds because of the energy extra requirements for acceleration at such velocity.
We have also developed a special relativistic version of the code but it requires much more numerical resources, and the results will be presented elsewhere.

\section{Results} \label{sec:results}

\subsection{Results of PIC simulations}\label{sec:picres}

\subsubsection{Shock structure}\label{sec:picres_sub1}

\begin{figure*}[!t]
\centering
\includegraphics[width=0.99\linewidth]{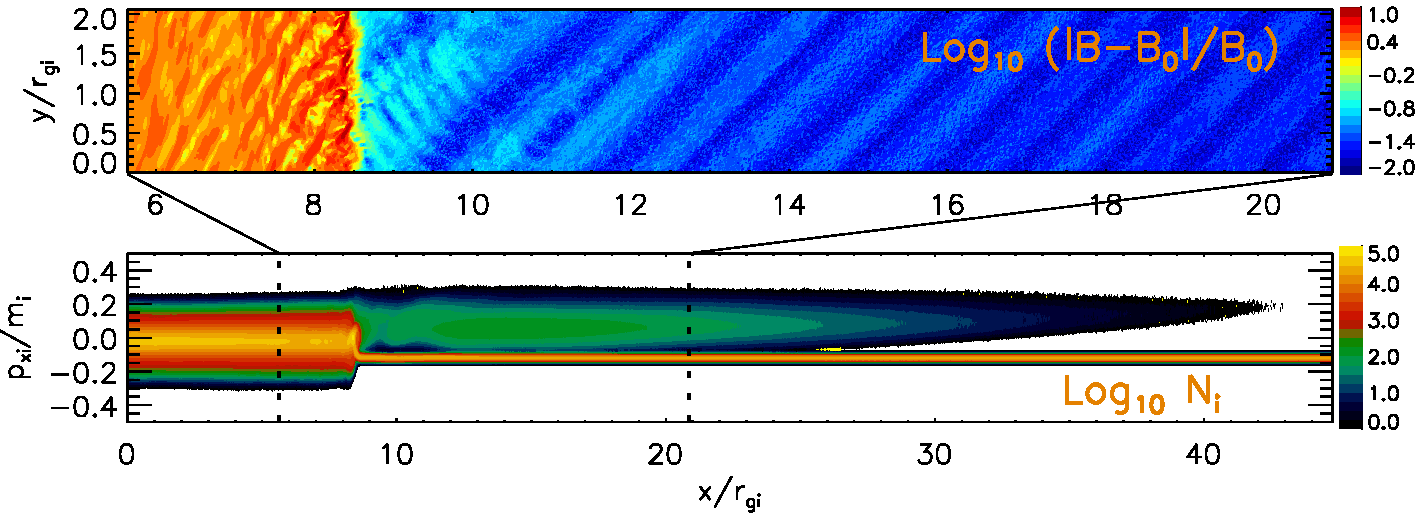}
\caption{Magnetic field map (top panel) and phase-space distribution of $p_{xi}$-$x$ (bottom panel) for run PIC15 at the last simulation step, $t \approx 68 \Omega_i^{-1}$.}
\label{pic_dens_ph}
\end{figure*}

Figure~\ref{pic_dens_ph} displays a \ab{magnetic field} map and the phase-space plot of $p_{xi}$-$x$ at the shock region of run PIC15. A portion of the upstream ions are accelerated via multiple SDA cycles and reflected back upstream \citep{2015ApJ...798L..28C}, where they can potentially be further accelerated via DSA. We henceforth denote these ions as injected. The fraction of injected ions is about $2\%$, which is in rough agreement with Caprioli's hybrid simulation \citep{2015ApJ...798L..28C} for the same shock parameters. 
The injected ions propagating upstream drive transverse waves with a wavelength of about $\lambda \approx 2 r_{\rm gi}$ (Fig.~\ref{pic_dens_ph}) and an amplitude of about $\delta B/B_0 \approx 0.1$ at $x = (9-10)r_{\rm gi}$, \mponn{consistent with the expected wavelength of resonant modes that are excited by ions streaming along the large-scale magnetic field with $v_\parallel \approx 0.3 c$.} 
\citet{Caprioli:2014b} \mponn{also} found upstream waves of resonant nature, corresponding to a wavelength $\lambda \lesssim 2\pi r_{\rm gi}$. However, we note that conditions at the shock upstream are also suitable for the non-resonant Bell mode \citep{2004MNRAS.353..550B,2005MNRAS.358..181B}. According to equations (16) and (17) from \cite{2006PPCF...48.1741R} the growth time and the wavelength of the fastest growing mode are $t_{\rm Bell} \approx 5 \omci^{-1}$ and $\lambda_{\rm Bell} \approx 3 r_{\rm gi}$ \ab{in the asymptotic limit}. 
Unfortunately, the simulation PIC15 is still too short to \ab{reach the steady state of the shock evolution and} say which mode is dominant in our case. 

This upstream turbulence is observed only in run PIC15. This run is characterized by the strongest injected ion flow, and the simulation time is long enough to disturb the upstream magnetic field. The DSA process, however, is still not initialized. Usually that requires a considerably longer simulation time, $T_\mathrm{sim} > 100 \Omega_i^{-1}$, hence the PIC-MHD simulations in our study (see Sec.~\ref{sec:picmhdres}).

\subsubsection{The ion injection efficiency}\label{sec:picres_sub2}

\begin{figure*}[!t]
\centering
\includegraphics[width=0.49\linewidth]{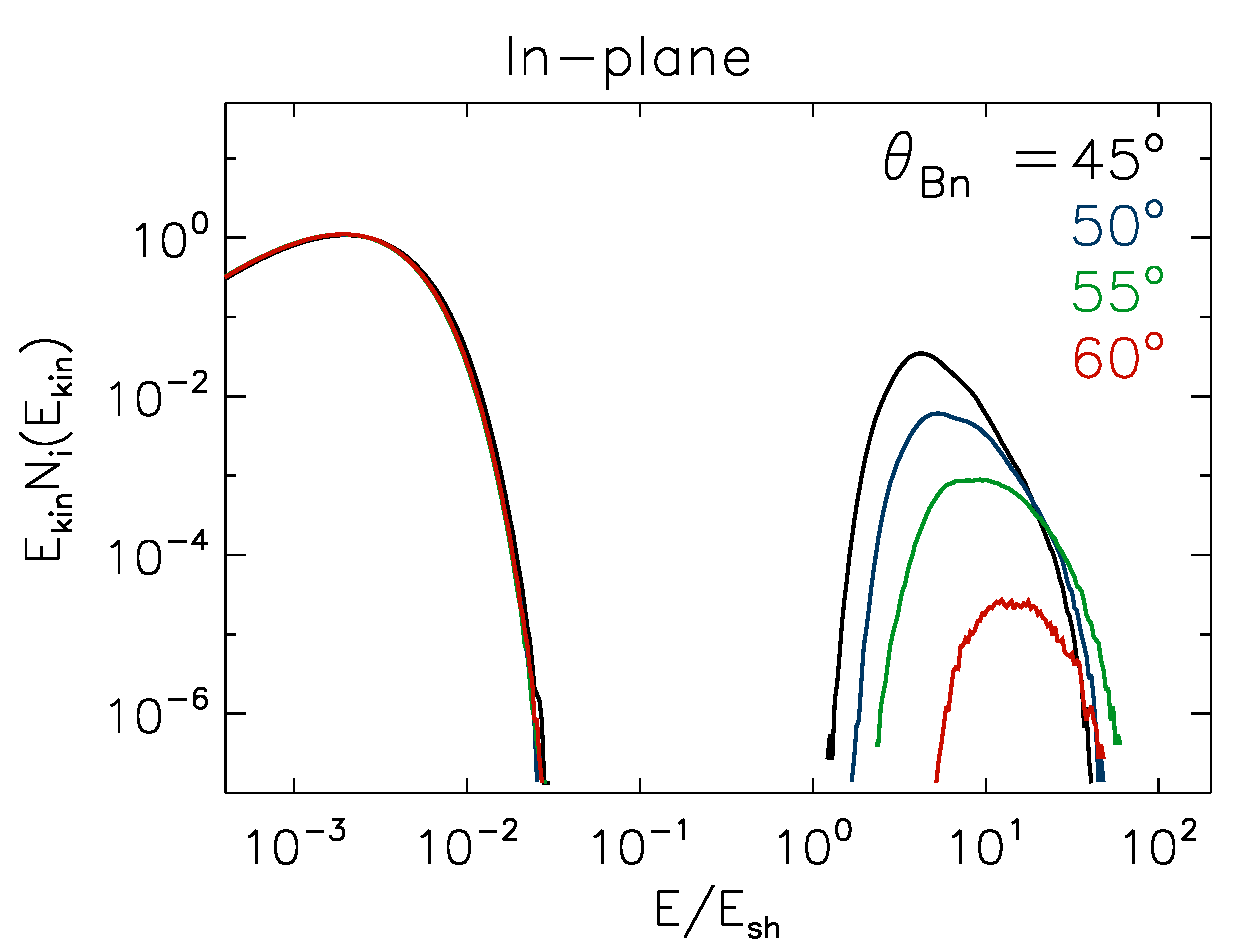}
\includegraphics[width=0.49\linewidth]{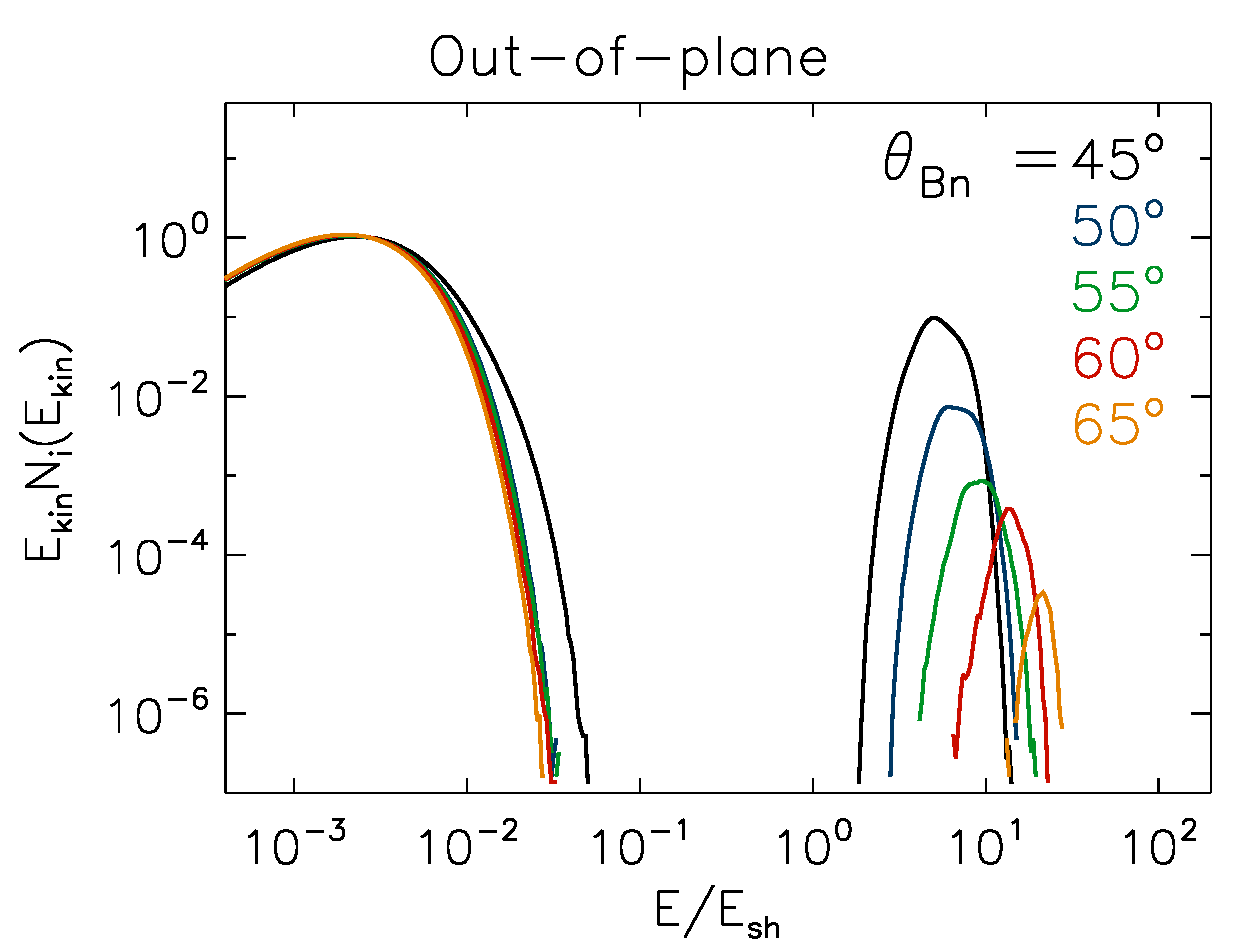}
\caption{The upstream ion spectra for runs with $\vsh/c = 0.133$, $\ma=20$, and $\beta_p= 1$. The left panel shows in-plane simulations, the right panel presents out-of-plane runs. Spectra are calculated for a region $x-x_{\rm sh} = (2-7)r_{\rm gi}$ ahead of the shocks and are averaged in time over at least $5\omci^{-1}$.}
\label{spectra_all_pic}
\end{figure*}

To study the ability to drive upstream turbulence, we use PIC simulations to determine the properties of injected ions as a function of the shock parameters. Table~\ref{table-param_pic} lists the density and energy fraction of the injected ions upstream of the shock. The energy fraction is defined as $U_{\rm inj}/U_{\rm sh}$, where $U_{\rm inj}$ is the energy density of injected ions in the upstream reference frame and $U_\mathrm{sh} = N_{\rm 0} \mi \vsh^2/2$. $U_{\rm inj}$ and $N_{\rm inj}$ are calculated over the region $x-x_{\rm sh} = (2-7)r_{\rm gi}$, where $x_{\rm sh}$ is the shock overshoot position, and averaged in time over at least over $5\omci^{-1}$. \ab{The fluctuation amplitude for $U_{\rm inj}$ and $N_{\rm inj}$ is at the level of $ 20\% - 30\%$. Note also that the injection efficiency remains stable in the longest run (PIC15), although waves are already evident upstream of the shock.}

The fraction of injected ions is mainly defined by the shock obliquity angle, $\theta_{B}$, and the angle $\varphi$.
Figure~\ref{spectra_all_pic} show the ion energy distributions in the upstream region for all simulations having $\vsh/c = 0.133$, $\ma=20$, and $\beta_p= 1$. The spectra are calculated in the upstream reference frame, and so the upstream ions are represented by Maxwell distributions predefined by the simulation setup. The injected ions have an energy above unity, on account of SDA. 
In all simulations the escaping ions have a similar average speed along the x-axis, which is about $1.6\vsh$ in the reference frame of the upstream plasma. To reach a such speed for higher $\theta_{B}$, the ions must have high velocity along the magnetic field and a high energy. In other words, at highly oblique shocks the ions must have experienced a large number of SDA cycles, which results in smaller injection rates. Shock simulations performed with out-of-plane magnetic field are usually characterized by larger fractions of injected ions compared to in-plane cases, because the overshoot is more coherent and provides better conditions for \ab{SDA}.

The ion injection efficiency is different for simulations with $\vsh/c=0.264$ and $\vsh/c=0.133$, the other parameters being the same (runs PIC03-PIC05 and PIC06-PIC08). In both cases the ions have approximately the same escape speed along the x-axis and it equals to $1.6\vsh$ if calculated in the upstream reference frame. Therefore, the total velocity is $v_\mathrm{inj}\simeq 1.6\vsh/\cos{\theta_B}=3.2\vsh$ and the average energy of ions propagating back upstream can be estimated as $\varepsilon_{\rm inj} \approx 12\,\varepsilon_{\rm sh}$ in slower shocks (runs PIC06-PIC08), while for faster shocks (runs PIC03-PIC05) it is $\varepsilon_{\rm inj} \approx 23\, \varepsilon_{\rm sh}$.
This difference is due to relativistic correction, therefore some extra energy ($+130\%$ for faster shocks and $+20\%$ for slower shocks) compared to the fully nonrelativistic case is needed to reach the same runaway speed. To achieve higher energies, ions should go through a larger number of SDA cycles, \mponn{and therefore fewer ions are} injected.
In fact, cases with $\vsh/c=0.133$ are very close to fully nonrelativistic shocks, as the \mponn{relativistic corrections are about $20\%$ which is commensurate with the variance} of the reflected ion fraction in PIC simulations. Therefore, \mponn{the} results can be directly applied for SNR shocks.

Shock simulations which differ only by the upstream plasma beta $\beta_{\rm p}$ \ab{(runs PIC08-PIC10) do not demonstrate a dependence on this parameter. The ion injection efficiency varies at the level just slightly above the random fluctuations while $\beta_{\rm p}$ is probed over two orders of magnitude.}

\ab{PIC simulations featuring shocks with different Alfv\'enic Mach numbers, $\ma$, demonstrate somewhat higher fluctuations of the ion injection efficiency. The results of simulations PIC06/PIC08 and PIC15/PIC16 are consistent within 30\% error bars. The injection efficiency for simulations PIC06 and PIC08 with higher shock velocity ($\vsh = 0.264c$) is a factor of 3 different. At such fast and oblique shocks the average injection velocity of ions is almost relativistic ($\sim 0.85c$) and even small fluctuations may result in a factor of a few different injection efficiencies. We also admit that when the ion injection is defined by SDA, the injection efficiency should not depend on $\ma$ (see also Sec.~\ref{sec:picres_sub3}).}

\ab{
The PIC simulation results suggest that the variable which affects the injection efficiency the most is the shock obliquity angle, $\theta_{B}$. Indeed, changing the shock obliquity angle from $\theta_{B}=45^o$ to $\theta_{B}=65^o$  leads to a sharp drop in the ion injection efficiency by five orders of magnitude, while other parameters change the injection efficiency by a factor of a few at most.}


\subsubsection{An ion injection model}\label{sec:picres_sub3}

\begin{figure}[!t]
\centering
\includegraphics[width=0.5\linewidth]{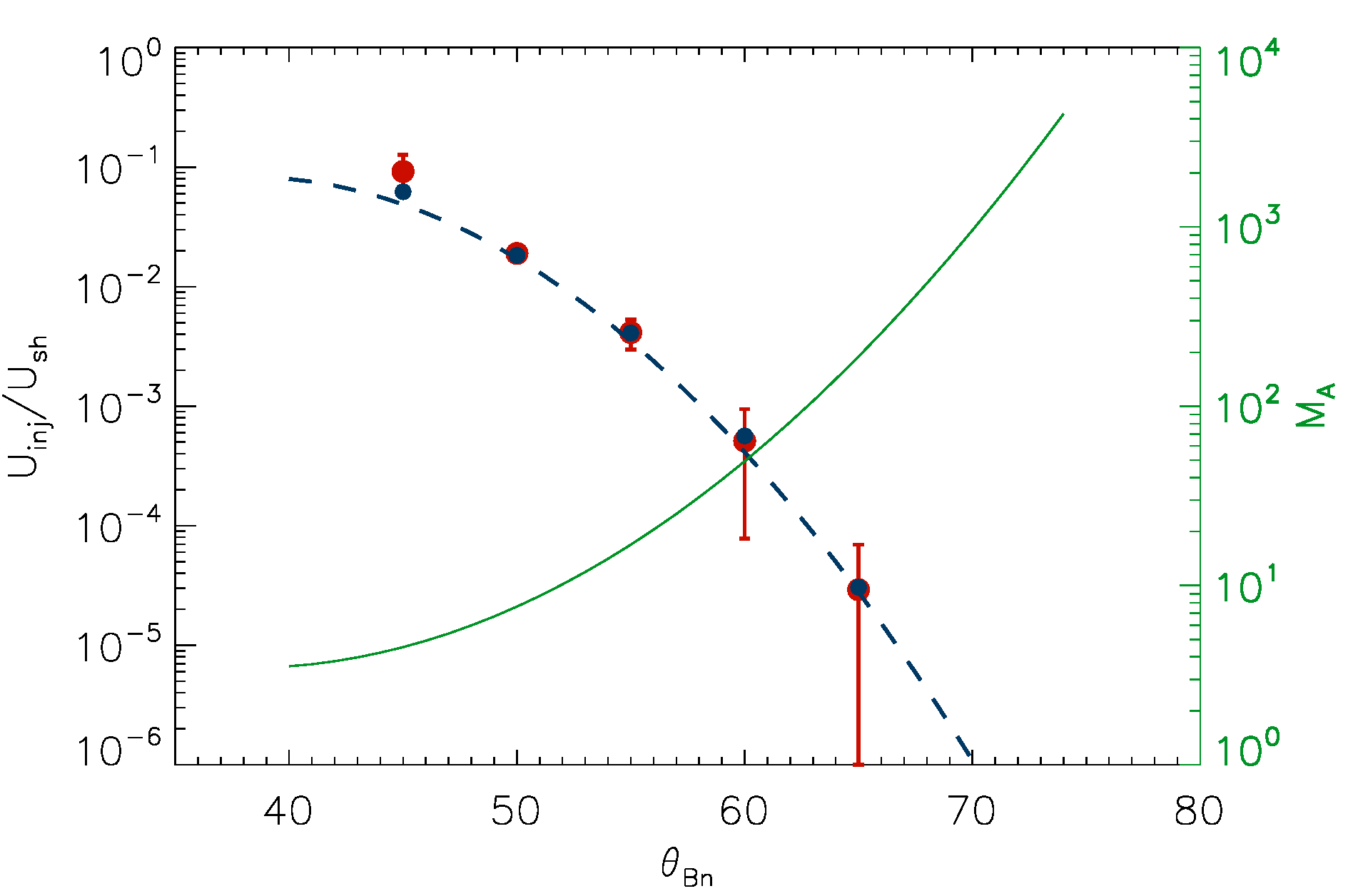}
\caption{The normalised energy density of injected ions in the upstream region: red dots are PIC simulation results \ab{(the error bars include both temporal variations and the difference between in-plane and out-of-plane runs)}, blue dots represent the ion injection model based on ion acceleration via multiple SDA cycles, and the dashed blue line is the best fit to the injection ion model, $F(\theta_{B})$ \ab{(Eq.~\ref{crit_mach})}. The green line gives the critical Alfv\'enic Mach number, $M_{\rm A,cr} = (F(\theta_{B}))^{-0.5}$, cf right axis.}
\label{frac_all_paper}
\end{figure}

As discussed in \citet{2015ApJ...798L..28C}, ions are accelerated through multiple SDA cycles and with sufficiently high velocity can escape the shock and become injected ions. During an SDA cycle an ion is reflected back upstream with conservation of the kinetic energy in the shock-rest frame.  
Then gyrating in the upstream region, the ion is accelerated by the motional electric field. The energy increment for cycle $n$ can be estimated as 
\be
\Delta \varepsilon_{\rm SDA,n} = e E_0 \Delta l_n
\ee
where $E_0 = v_{\rm sh} B_0 \sin{\theta_{B}}$, $\Delta l_n = \alpha r_{g,n} = \alpha p_{i,n}/(e B_0) $ and $\alpha$ defines the acceleration efficiency of an individual SDA cycle.
Therefore the ion kinetic energy after $(n+1)$ SDA cycles can be defined as
\be
\varepsilon_{\rm SDA,n+1} = \varepsilon_{\rm SDA,n} + \alpha c v_{\rm sh} \mi \sqrt{\gamma_n^2-1} \sin{\theta_{B}} \ .
\label{ene_one_cycl}
\ee
For each $\theta_{B}$ we calculate how many SDA cycles are needed to achieve the escape speed of $v_{\rm x}=1.6 v_{\rm sh}$ \ab{in the upstream reference frame}. When the escape speed is fixed the ion injection efficiency is defined by the balance between the escape probability $p$ and the acceleration efficiency $\alpha$. The fraction of ions escaping downstream after each SDA cycle was derived by \cite{2015ApJ...798L..28C} \mponn{as} $p=0.75$. Therefore $\alpha$ \mponn{must be equal to 0.34 for the injection model to match the} simulation results.
The thermal energy of injected ions is calculated assuming that the x-projection of the thermal velocity is comparable to the velocity difference between the shock and escaping particles, namely, $v_{\rm th} \approx 0.6 v_{\rm sh}/\cos{\theta_{B}}$. The resulting energy density fractions are marked by blue dotes in Figure~\ref{frac_all_paper}. The dashed line is a fit, 
\be
\frac{U_{\rm inj}}{U_{\rm sh}} = F(\theta_{B}) = 10^{a_0+a_1\cdot\theta_{B}+a_2\cdot\theta_{B}^2}\ ,
\label{crit_mach}
\ee
where $a_0 = -7.3$, $a_1= 0.35$ and $a_2= -0.0047$. Note, that the escape speed in PIC simulations slightly depends on the shock obliquity, it increases from $1.55 v_{\rm sh}$ to $1.69 v_{\rm sh}$, when the shock obliquity changes from $\theta_{B}=45^o$ to $\theta_{B}=65^o$. Hence the escape probability or SDA acceleration efficiency might also slightly depend on the shock obliquity. Taking into account that the final results only weakly depend on these vagaries, here for simplicity we assume that all these parameters are constant. 

The red dots in Figure~\ref{frac_all_paper} shows results for runs with $\vsh/c = 0.133$, $\ma=20$ and $\beta_p= 1$. Each point represents the average for in-plane and out-of-plane configurations for the $\theta_{B}$ in question. The error bars include both temporal variations and the difference between in-plane and out-of-plane runs. Figure~\ref{frac_all_paper} demonstrates a good match between PIC simulation results and the injection model. The green line represents the critical Mach number, $M_{\rm A,cr} = (F(\theta_{B}))^{-0.5}$, at which the energy density of injected ions equals that of the magnetic field in the upstream region, and the injected ions \secrev{are able to drive a considerably stronger upstream turbulence necessary for efficient DSA.}

\ab{ For all shocks with $\ma \gg 20 $ we shall assume that the ion injection rate does not depend on the Alfv\'enic Mach number, because the ion injection process is driven by the shock potential which is the same (when it is normalized to the upstream bulk energy) for all high-Mach-number shocks. Also, \mponn{the energy gain} during one SDA cycle does not depend on the Alfv\'enic Mach number (see Eq.~\ref{ene_one_cycl}). At perpendicular high-Mach-number shocks, the physics of which is very similar to \mponn{that of} oblique shocks, the ion behaviour does not depend on $\ma$ for a wide range of Mach numbers, $\ma \approx 20-70$ in \cite{Bohdan2020a} and $\ma \approx 130$ in \cite{2010ApJ...721..828K}. }

\ab{ 
Summarizing, we can state that as long as SDA works in the usual regime, the ion acceleration efficiency should remain roughly constant and the injection rate is largely defined by $\theta_{B}$, while the influence of the plasma beta of Alfv\'enic Mach number is rather minor, if existent at all. Therefore, the best fit to the injection model (the dashed blue line in Fig.~\ref{frac_all_paper}) can be used to represent the ion injection energy fraction in MHD-PIC simulation.}

\subsection{Results of PIC-MHD simulations}\label{sec:picmhdres}

\begin{figure}[!tb]
\centering
\includegraphics[width=0.99\columnwidth]{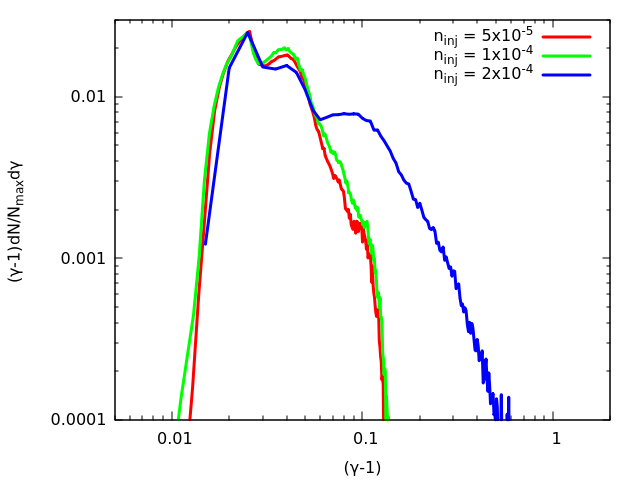}
\caption{SEDs for simulations PICMHD01B, -02 and -03 with $\ma=20$ and varying injection rates at time \ab{$t=3000\,\Omega_{\rm inj}^{-1}$}. At the lower injection rates, we see only \mponn{the bump at $\gamma-1=0.04$ that is caused by} shock-drift acceleration, \ab{For the highest injection rate, we see a secondary bump at $\gamma-1=0.1$ \mponn{populated by particles that experienced} a second round of acceleration. At higher energies, the SED shows a constant slope, indicating that at least some particles \mponn{have been further accelerated}. This marks the very beginning of DSA.} }
\label{fig:SED_M20}
\end{figure}

\begin{figure}[!tb]
\centering
\includegraphics[width=0.99\columnwidth]{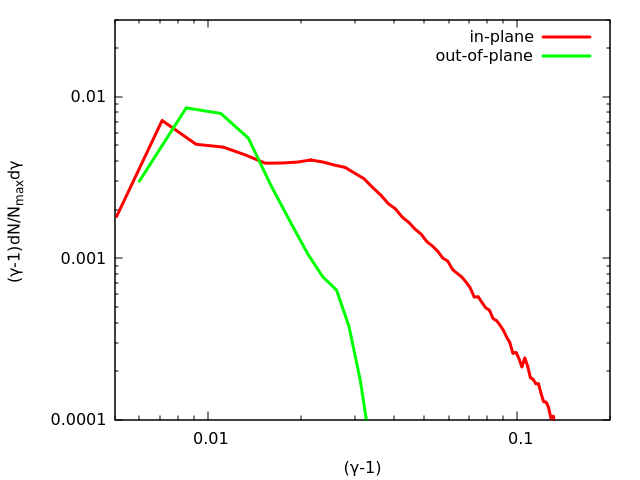}
\caption{SEDs for simulations PICMHD05A and -05B at time \ab{$t=3000\,\Omega_{\rm inj}^{-1}$}, distinguishing the difference between a magnetic field in and out of the plane of the simulation. Clearly, DSA does not take place when the magnetic field is out-of-plane.}
\label{fig:SED_ip_oop}
\end{figure}

\begin{figure}[!tb]
\centering
\includegraphics[width=0.99\columnwidth]{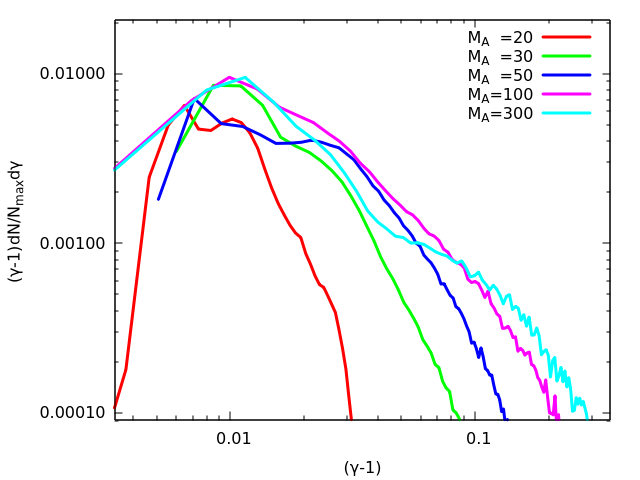}
\caption{SEDs for simulations with $\beta_{\rm p}$=1 and varying Alfv{\'e}nic Mach numbers at time \ab{$t=3000\,\Omega_{\rm inj}^{-1}$}. At $\ma=20$, we see no significant acceleration through DSA. At $\ma=30$ a high energy tail starts to develop, indicating DSA, similar to what is observed for $\ma=20$ with an exaggerated injection rate. This feature becomes more prominent at higher Alfv{\'e}nic Mach numbers.}
\label{fig:SED_mach}
\end{figure}

\begin{figure}[!tb]
\centering
\includegraphics[width=0.99\columnwidth]{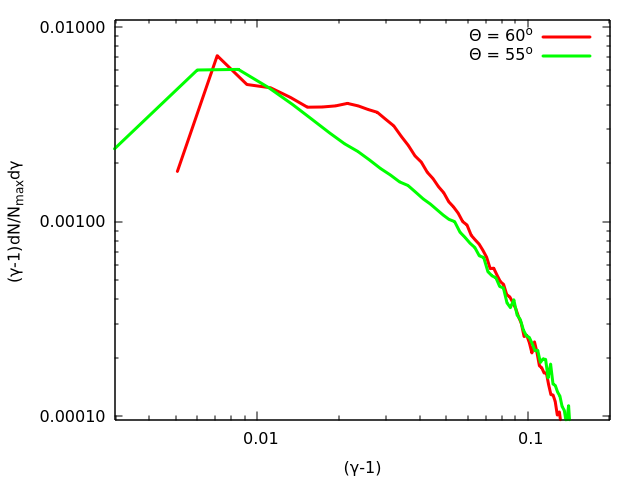}
\caption{SEDs at time \ab{$t=3000\,\Omega_{\rm inj}^{-1}$} for simulations PICMHD05A and -12 with  $\theta_B=60$ and 55 degrees. Both show clear evidence of DSA, but the 55-degree model lacks the "bump" created by shock drift acceleration.}
\label{fig:SEDbangle}
\end{figure}

\begin{figure}[!tb]
\centering
\includegraphics[width=0.99\columnwidth]{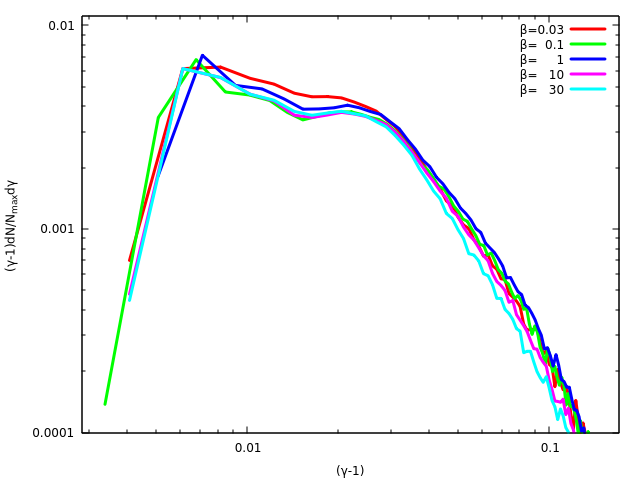}
\caption{SEDs for different plasma beta with $\ma=50$ at time \ab{$t=3000\,\Omega_{\rm inj}^{-1}$}. All SEDs are very close together, 
indicating that $\beta_{\rm p}$ has little or no influence on the SED for a given Alfv{\'e}nic Mach number.}
\label{fig:SEDbeta}
\end{figure}

\subsubsection{Low Mach numbers and $\beta_p$=1 simulations}
The results obtained in section \ref{sec:picres} show that the injection rate strongly depends on the angle between the magnetic field and the flow, $\theta_{B}$, and this in turn directly influences the evolution of the upstream medium. We have run a series of simulations with an obliquity angle $\theta_{\rm B}=60^o$ and with Alfv\'enic Mach number $\ma\,=\,20$ (PICMHD01A-03), based on the PIC results to test at which injection rate DSA starts to occur. The results of these simulations are presented in Figure~\ref{fig:SED_M20}, which shows the spectral energy distribution (SED) of the non-thermal particles \ab{at $t=3000\,\Omega_{\rm inj}^{-1}$ with $\Omega_{\rm inj}$ in gyrofrequency at injection defined by the upstream magnetic  field $B_0$ and the injection velocity $v_{\rm inj}$. 
For simulations with an injection rate of $
5\times10^{-5}$ we see no \mponn{spectral tail with constant slope, and hence no} significant DSA. Some of the particles are accelerated, but this is due to the SDA process as was demonstrated in \citet{vanMarle:2018} and discussed in the previous section.}
\ab{This process creates the ``bump'' at $\gamma-1\,=\,0.04$. The simulation with an injection rate of $\mponn{2}\times10^{-4}$ does show an extension toward higher energies. Its most prominent feature is a secondary bump at $\gamma-1\,=\,0.1$, indicating that some of the particles that escape from the shock are reflected back toward the shock to \mponn{go through} a second round of shock-drift acceleration. Furthermore, instead of falling off immediately, this SED shows a nearly constant slope at higher energies, indicating that at least some DSA occurs. The slope has an index that corresponds with $dN/d\gamma \propto (\gamma-1)^{-3.4}$. This is much steeper than expected for a typical high-Mach shock (approximately -1.5). However, the SDA process remains incomplete due to the low amplitude of the upstream instabilities. As a result, many particles escape without being reflected, which leads to a steeper spectrum.}
However, we find that this required injection rate is higher than what is obtained through PIC simulations. Therefore we can conclude that at $\ma=20$, DSA \mponn{does likely} not occur in shocks with an obliquity of 60 degrees or more.

\subsubsection{High Alfv\'enic Mach number runs}
\label{sec-highMA}
In a second series of runs (PICMHD04-07) we inject flows at high Alfv\'enic Mach numbers ($\ma\ge\,30$). This regime is not accessible to PIC simulations (at least for a reasonable computational expense). 
We do this by changing the magnetic field strength. The  other parameters remain the same, including the injection rate, which is fixed at $5\times10^{-5}$, based on the results of PIC07 and PIC08, because the PIC simulations showed that there is no strong dependence of the injection rate on the Mach number. This prediction is also born out by \citet[][Fig.~3]{Caprioli:2014a}, who argue that at high obliquity the injection rate becomes nearly independent of the shock Mach number. PIC models by \citet{2018ApJ...864..105H} show an increase in efficiency for increased Mach numbers, but this was for low-Mach-number shocks ($M_s\,<\,3.5$) for which the compression ratio increases with the Mach number. This is not the case in our models, which all pertain to high-Mach-number shocks.

As is shown in Fig~\ref{fig:SED_mach}, the shape of the SED changes strongly between $\ma=20$ and $\ma=30$. The spectrum develops a high-energy tail that extends further for higher Mach numbers. This matches the prediction obtained from Fig.~\ref{frac_all_paper} that for Alfv{\'e}nic Mach numbers higher than the critical one (for an obliquity angle of 60$^o$ the critical Alfv\'enic Mach number is around 30), the injection rate would be sufficient to trigger instabilities in the upstream medium.

For comparison we include a simulation (PICMHD05B) with magnetic field going out of the plane of the simulation, rather than lying in the plane. Fig.~\ref{fig:SED_ip_oop} demonstrates that the simulation with the magnetic field out of the 2D plane shows no sign of DSA. Under these circumstances, the particles cannot trigger an instability since their direction of motion is perpendicular to the plane of the simulation. 

\subsubsection{55 vs. 60 degrees}
\ab{We compare the results for shocks with obliquity of 55 and 60 degrees (PICMHD05A and PICMHD12) in Figure~\ref{fig:SEDbangle} to demonstrate the transition between shocks that still follow the general model of the quasi-parallel shocks and the truly oblique shocks}. The lower obliquity corresponds to a much higher injection rate (See simulations PIC11 and PIC12), for which the issue of DSA is not in doubt. However, the two spectra show a marked difference. The shock with 55 degrees obliquity lacks the "bump" at $(\gamma-1)\,=\,0.03$, which is clearly visible at 60 degrees. This bump is created by shock-drift acceleration, which is less effective at lower obliquity. Instead, the SED shows only DSA from $(\gamma-1)\,=\,0.01-0.05$. 
At these injection rates, the velocity and Alfv{\'e}nic Mach number of the shock become largely irrelevant because the injected particle density is sufficient to trigger DSA for all non-relativistic shocks.

\subsubsection{Investigation of plasma beta effects}
The effect of varying plasma beta is demonstrated in Fig.~\ref{fig:SEDbeta}, which shows the SEDs for simulations PICMHD08-11 compared to the model with the same Alfv{\'e}nic Mach number and $\beta_{\rm p}$=1 (PICMHD05A). 
Here we vary the plasma beta by changing the thermal pressure and keeping the magnetic field constant. Again, as in Sect.~\ref{sec-highMA}, we do not vary the injection rate. 
PIC simulations of shocks with high plasma-$\beta$ show that these shocks have no problem producing non-thermal particles as long as the sonic Mach number remains above 3.5 \citep{2018ApJ...864..105H}. In our models we do not reach this point and all shocks can be considered to be strong.

Clearly, the variation in $\beta_{\rm p}$ has little influence on the SED, although there is a slight trend to have less efficient DSA for higher $\beta_{\rm p}$. This trend can be explained by the fact that at higher plasma beta the thermal pressure tends to counteract variations in the magnetic field, thereby reducing the local distortion. This effect was demonstrated for extreme high-$\beta_{\rm p}$ shocks by \citet{vanMarle2019b,Vanmarle:2020}.

\subsubsection{\secrev{Resonant vs. non-resonant instability}}
As already stated in section 3.1.1 shocks with low Alfvénic Mach number can also trigger the resonant streaming instability over short timescales. This branch has no restriction over the magnetic energy density of the injected ions with respect to the background magnetic energy. Following the discussion in \cite{Caprioli:2014b} (their Eq.9) the ratio of the resonant to the non-resonant growth rate can be written as:
\begin{equation}
    \frac{\Gamma_{\rm res}}{\Gamma_{\rm NRH}} \simeq \frac{1}{\sqrt{N_{\rm inj}/N_0}} \frac{1}{M_{\rm A}}
\end{equation}
To derive this expression we have considered that the injection fraction scales as $1/p$.  \secrev{For numerically manageable injection rates, the resonant branch dominates over the non-resonant one only at low Alfv\'enic Mach numbers, for which the standard formula for the non-resonant growth rate violates the condition $k r_\mathrm{L}\gg 1$. In that case, the resonant modes should be seen, if $N_{\rm inj}/N_0$ is large enough to provide a few growth cycles upstream of the shock. 
At high Alfv\'enic Mach numbers the resonant mode is dominant only for very low $N_{\rm inj}/N_0$ ratios. This does not mean that DSA cannot be supported by the resonant instability in the high-Mach-number regime, but that it would require too long a simulation to be demonstrated. The lower $N_{\rm inj}/N_0$, the lower are the growth rates and the saturation field strength for both branches of the instability, and the longer is the acceleration timescale \citep[Eq.8 in][]{Caprioli:2014b}. 
}

\secrev{Hence, considering in parallel the critical Alfv\'enic Mach constraint in Section~\ref{sec:picres_sub3} it is difficult at high shock obliquities to isolate a parameter subspace where the resonant unambiguously dominates the magnetic turbulence spectrum especially close or even below the critical Alfv\'enic Mach number. This issue deserves a dedicated future study.}

\section{Discussion} \label{sec:discussion}

PIC-MHD simulations by \citet{vanMarle:2018} showed that DSA could occur at oblique shocks, provided that the injection rate of non-thermal particles was sufficiently large, and that the energy density of the upstream current exceeded that of the local magnetic field. However, as our PIC simulations show, the injection rate is strongly dependent on the angle between the flow and the magnetic field, with injection rates dropping rapidly for shocks with $\theta_{B_n}>50$. 
\ab{They also show that the injection is, fundamentally, a 3D process, as demonstrated by the difference in injection rates between in-plane and out-of-plane simulations.}

We find that for $\theta_{B_n}=60$ the injection rate is insufficient for shocks at $M_{\rm A}=20$. However, for shocks with a higher Alfv{\'e}nic Mach number, the energy density of the injected upstream particles is still sufficient to trigger local instabilities in the magnetic field, thereby producing DSA. 
In contrast to the strong influence of the Alfv{\'e}nic Mach number, the plasma beta of the upstream medium proves to be mostly irrelevant, \mponn{provided} that all shocks are still effectively high-Mach\mponn{-number} shocks. 

We can conclude that high-Mach-number oblique shocks are capable of accelerating particles through DSA, whereas for slower shocks acceleration becomes impossible at angles of more than approximately 50-55 degrees, which matches previous results \citep[e.g.][]{Caprioli:2014a, Caprioli:2014b, Caprioli:2014c,Haggerty:2019}. 

Shocks with high Alfv{\'e}nic Mach numbers are quite common in astrophysical objects. Stellar winds can reach velocities of several thousands kilometers per second \citep[e.g.][]{2011A&A...534A..97K}, which, compared to the typically low magnetic field strength in both the wind bubble and the interstellar medium can lead to flows with high Alfv{\'e}nic Mach numbers, as demonstrated by \citet{2020MNRAS.493.4172S}. Even higher velocities can be found in expanding supernova remnants (SNRs). Considering the results obtained in this article we also expect ions to be injected at large obliquities in SNRs like SN 1006. 
This object  has shock speeds of the order of 3000\,km/s (see \citet{parizot_2006:snr_bampli} and references therein) or even up to 5800\, km/s \citep{Ressler2014}. Adopting a background magnetic field of 3$\mu$G and a gas density of 0.05 $\rm{cm}^{-3}$ \citep{Acero2007}, we find that SN 1006 shocks have Alfv\'enic Mach numbers $\sim 100-200$ which would allow ion injection up to obliquities $\sim 65^o$. These finding raise doubts that the bipolar morphology of SN~1006 can be easily explained as the signature of a relatively homogeneous magnetic field in the environment \citep{2011A&A...531A.129B}. \ab{Finally, results obtained by \citet{2016ApJ...827...36O} that  shocks propagating into a partially ionized plasma demonstrate an increased injection rate, allowing for DSA in quasi-perpendicular shocks. This condition applies to many astrophysical shocks, such as a supernova remnant expanding into cold interstellar medium.}

\citet{2017ApJ...840..112S} found velocities of up to 7800\,km/s in \object{Tycho}'s SNR. But the shock \mponn{speed seems to be} rather position-dependent \citep[and references therein]{Morlino2016}, as in its northeast part the remnant appears to interact with gas denser than typical, 0.2-1 ${\rm cm}^{-3}$ \citep{Williams2013}. Accounting for these variations and again assuming a background magnetic field strength of 3 $\mu$G, we find typical Alfv\'enic Mach numbers between 200 and 1000, for which obliquities above $65^o$ may allow ion injection. 

At the initial supernova break-out the shock speed can be higher by up to an order of magnitude. \citet[][page 6 and 27]{2017hsn..book..967W} predict break-out speeds of $10^9-10^{10}$~cm/s for core-collapse supernovae of red and blue supergiants. Such ejecta would interact with the moderately magnetized free stellar wind, leading to shocks with high Alfv{\'e}nic Mach numbers up to a few hundreds to one thousand. CR-driven instabilities are expected to grow over day timescales, leading to fast particle acceleration \citep{Marcowith2018, 2021arXiv210813433I}. In these configurations, again, highly oblique shocks may allow ion injection.

Finally once turbulence is triggered by ions, electrons having sufficiently high energies can also be accelerated via DSA, giving rise of synchrotron radiation in radio and X-rays. Indeed, electrons at oblique quasi-perpendicular shocks can be accelerated up to relativistic energies \citep{Matsumoto2017,Xu2020} injecting them into DSA via the shock internal mechanisms, such as, electron shock-surfing acceleration \citep{Matsumoto2012,Bohdan2019a}, magnetic reconnection \citep{Matsumoto2015,Bohdan2020a}, SDA \citep{Xu2020}, and stochastic Fermi-like acceleration \citep{Matsumoto2017,Bohdan2017}. However, the exact fraction of injected electrons still should be defined with further PIC numerical experiments.

\section{Conclusions} \label{sec:conclusions}
\mponn{Combining PIC and PIC-MHD techniques, we investigate the efficiency of ion into the DSA process as a} function of the Alfv\'enic Mach number, $\ma$, the magnetic field obliquity, $\theta_{\rm B}$, and the plasma $\beta$ parameter. Our main findings are:

\begin{itemize}
    \item The ion injection efficiency in PIC simulations of oblique non-relativistic shocks is \ab{largely} defined by the shock obliquity angle, $\theta_{\rm B}$. \mponn{The influence of the sonic and the Alfv\'enic Mach number as well as} the upstream plasma beta is \ab{minor and} commensurate with measurement errors. The ion energy injection efficiency can be described with the equation $U_{\rm inj}/U_{\rm sh} = F(\theta_{\rm B})= 10^{a_0+a_1\cdot\theta_{B}+a_2\cdot\theta_{B}^2}$, where $a_0$=-7.3, $a_1=$0.35, and $a_2=-0.0047$. 
    \item Shocks do not permit the triggering of DSA at a fix obliquity if the shock Alfv\'enic number is below the empirical critical value $M_{\rm cr} = F(\theta_{\rm B})^{-0.5}$, at which the \mponn{energy density of injected ions in the upstream region is equal to that of the background magnetic field}. For example, at oblique shocks with $\theta_{\rm B} = 60^o$ DSA can be already triggered if $M_{\rm A}  \gtrsim 46$. 
    \item  \ab{In the case of the $\ma=\,50$ shock,} the influence of plasma-$\beta$ on both the injection efficiency and the acceleration is weak. \mponn{However, this will cease to be the case if plasma-$\beta$ is high enough that the sonic Mach number becomes smaller than $\ms\,\simeq\,3.5$, as demonstrated by \citet{2018ApJ...864..105H} and \citet{Vanmarle:2020}.}
\end{itemize}

High Alfv{\'e}nic Mach numbers in excess to 100, which are routinely found in the interstellar medium, can hence trigger \secrev{efficient} DSA up to rather high magnetic obliquity of the order of $60^o$. However, at higher obliquity, the critical energy criterion quickly becomes impossible to fulfill as it would require Alfv{\'e}nic Mach numbers of a thousand or more, which are unlikely to occur for most astrophysical shocks. \secrev{Our study does not preclude that the resonant branch of the streaming instability may control the turbulence development at low Alfv\'enic Mach numbers close to or even below the critical value.}

\begin{acknowledgments}
PIC numerical experiments were conducted on resources provided by the North-German Supercomputing Alliance (HLRN) under the project bbp00033 and by the Prometheus system at Academic Computer Centre Cyfronet AGH. This work is supported by the  ANR-19-CE31-0014GAMALO project. PIC-MHD simulations were performed on the OCCIGEN machine at CINES under project: A0100412387
\end{acknowledgments}

\bibliographystyle{apj}
\bibliography{ref}

\begin{thebibliography}{49}
\expandafter\ifx\csname natexlab\endcsname\relax\def\natexlab#1{#1}\fi

\bibitem[{{Acero} {et~al.}(2007){Acero}, {Ballet}, \&
  {Decourchelle}}]{Acero2007}
{Acero}, F., {Ballet}, J., \& {Decourchelle}, A. 2007, \aap, 475, 883

\bibitem[{{Bell}(1978)}]{Bell:1978}
{Bell}, A.~R. 1978, \mnras, 182, 147

\bibitem[{{Bell}(2004{\natexlab{a}})}]{Bell:2004}
---. 2004{\natexlab{a}}, \mnras, 353, 550

\bibitem[{{Bell}(2004{\natexlab{b}})}]{2004MNRAS.353..550B}
---. 2004{\natexlab{b}}, \mnras, 353, 550

\bibitem[{{Bell}(2005)}]{2005MNRAS.358..181B}
---. 2005, \mnras, 358, 181

\bibitem[{{Blandford} \& {Ostriker}(1978)}]{Blandford:1978}
{Blandford}, R.~D., \& {Ostriker}, J.~P. 1978, \apjl, 221, L29

\bibitem[{{Bocchino} {et~al.}(2011){Bocchino}, {Orlando}, {Miceli}, \&
  {Petruk}}]{2011A&A...531A.129B}
{Bocchino}, F., {Orlando}, S., {Miceli}, M., \& {Petruk}, O. 2011, \aap, 531,
  A129

\bibitem[{{Bohdan} {et~al.}(2017){Bohdan}, {Niemiec}, {Kobzar}, \&
  {Pohl}}]{Bohdan2017}
{Bohdan}, A., {Niemiec}, J., {Kobzar}, O., \& {Pohl}, M. 2017, \apj, 847, 71

\bibitem[{{Bohdan} {et~al.}(2019){Bohdan}, {Niemiec}, {Pohl}, {Matsumoto},
  {Amano}, \& {Hoshino}}]{Bohdan2019a}
{Bohdan}, A., {Niemiec}, J., {Pohl}, M., {Matsumoto}, Y., {Amano}, T., \&
  {Hoshino}, M. 2019, \apj, 878, 5

\bibitem[{{Bohdan} {et~al.}(2020){Bohdan}, {Pohl}, {Niemiec}, {Vafin},
  {Matsumoto}, {Amano}, \& {Hoshino}}]{Bohdan2020a}
{Bohdan}, A., {Pohl}, M., {Niemiec}, J., {Vafin}, S., {Matsumoto}, Y., {Amano},
  T., \& {Hoshino}, M. 2020, \apj, 893, 6

\bibitem[{{Buneman}(1993)}]{Buneman1993}
{Buneman}, O. 1993, Computer Space Plasma Physics: Simulation Techniques and
  Software Eds.: H. Matsumoto \& Y. Omura, Tokyo: Terra Scientific, 67

\bibitem[{{Caprioli} {et~al.}(2015){Caprioli}, {Pop}, \&
  {Spitkovsky}}]{2015ApJ...798L..28C}
{Caprioli}, D., {Pop}, A.-R., \& {Spitkovsky}, A. 2015, \apjl, 798, L28

\bibitem[{{Caprioli} \& {Spitkovsky}(2014{\natexlab{a}})}]{Caprioli:2014a}
{Caprioli}, D., \& {Spitkovsky}, A. 2014{\natexlab{a}}, \apj, 783, 91

\bibitem[{{Caprioli} \& {Spitkovsky}(2014{\natexlab{b}})}]{Caprioli:2014b}
---. 2014{\natexlab{b}}, \apj, 794, 46

\bibitem[{{Caprioli} \& {Spitkovsky}(2014{\natexlab{c}})}]{Caprioli:2014c}
---. 2014{\natexlab{c}}, \apj, 794, 47

\bibitem[{Dorobisz {et~al.}(2018)Dorobisz, Kotwica, Niemiec, Kobzar, Bohdan, \&
  Wiatr}]{10.1007/978-3-319-78024-5_15}
Dorobisz, A., Kotwica, M., Niemiec, J., Kobzar, O., Bohdan, A., \& Wiatr, K.
  2018, in Parallel Processing and Applied Mathematics, ed. R.~Wyrzykowski,
  J.~Dongarra, E.~Deelman, \& K.~Karczewski (Cham: Springer International
  Publishing), 156--165

\bibitem[{{Drury}(1983)}]{Drury:1983}
{Drury}, L.~O. 1983, Reports on Progress in Physics, 46, 973

\bibitem[{{Friedman}(1990)}]{Friedman1990}
{Friedman}, A. 1990, US–Japan Workshop on Advanced Computer Simulation
  Techniques Applied to Plasmas and Fusion

\bibitem[{{Ha} {et~al.}(2018){Ha}, {Ryu}, {Kang}, \& {van
  Marle}}]{2018ApJ...864..105H}
{Ha}, J.-H., {Ryu}, D., {Kang}, H., \& {van Marle}, A.~J. 2018, \apj, 864, 105

\bibitem[{{Haggerty} \& {Caprioli}(2019)}]{Haggerty:2019}
{Haggerty}, C.~C., \& {Caprioli}, D. 2019, \apj, 887, 165

\bibitem[{{Hanusch} {et~al.}(2019){Hanusch}, {Liseykina}, {Malkov}, \&
  {Aharonian}}]{Hanusch2019}
{Hanusch}, A., {Liseykina}, T.~V., {Malkov}, M., \& {Aharonian}, F. 2019, \apj,
  885, 11

\bibitem[{{Inoue} {et~al.}(2021){Inoue}, {Marcowith}, {Giacinti}, {van Marle},
  \& {Nishino}}]{2021arXiv210813433I}
{Inoue}, T., {Marcowith}, A., {Giacinti}, G., {van Marle}, A.~J., \& {Nishino},
  S. 2021, arXiv e-prints, arXiv:2108.13433

\bibitem[{{Kato} \& {Takabe}(2010)}]{2010ApJ...721..828K}
{Kato}, T.~N., \& {Takabe}, H. 2010, \apj, 721, 828

\bibitem[{{Krti{\v{c}}ka} \& {Kub{\'a}t}(2011)}]{2011A&A...534A..97K}
{Krti{\v{c}}ka}, J., \& {Kub{\'a}t}, J. 2011, \aap, 534, A97

\bibitem[{{Kumar} \& {Reville}(2021)}]{Kumar2021}
{Kumar}, N., \& {Reville}, B. 2021, \apjl, 921, L14

\bibitem[{{Marcowith} {et~al.}(2018){Marcowith}, {Dwarkadas}, {Renaud},
  {Tatischeff}, \& {Giacinti}}]{Marcowith2018}
{Marcowith}, A., {Dwarkadas}, V.~V., {Renaud}, M., {Tatischeff}, V., \&
  {Giacinti}, G. 2018, \mnras, 479, 4470

\bibitem[{{Marcowith} {et~al.}(2020){Marcowith}, {Ferrand}, {Grech}, {Meliani},
  {Plotnikov}, \& {Walder}}]{Marcowith:2020}
{Marcowith}, A., {Ferrand}, G., {Grech}, M., {Meliani}, Z., {Plotnikov}, I., \&
  {Walder}, R. 2020, Living Reviews in Computational Astrophysics, 6, 1

\bibitem[{{Matsumoto} {et~al.}(2012){Matsumoto}, {Amano}, \&
  {Hoshino}}]{Matsumoto2012}
{Matsumoto}, Y., {Amano}, T., \& {Hoshino}, M. 2012, \apj, 755, 109

\bibitem[{{Matsumoto} {et~al.}(2015){Matsumoto}, {Amano}, {Kato}, \&
  {Hoshino}}]{Matsumoto2015}
{Matsumoto}, Y., {Amano}, T., {Kato}, T.~N., \& {Hoshino}, M. 2015, Science,
  347, 974

\bibitem[{{Matsumoto} {et~al.}(2017){Matsumoto}, {Amano}, {Kato}, \&
  {Hoshino}}]{Matsumoto2017}
---. 2017, Phys. Rev. Lett.

\bibitem[{{Morlino} \& {Blasi}(2016)}]{Morlino2016}
{Morlino}, G., \& {Blasi}, P. 2016, \aap, 589, A7

\bibitem[{{Niemiec} {et~al.}(2008){Niemiec}, {Pohl}, {Stroman}, \&
  {Nishikawa}}]{2008ApJ...684.1174N}
{Niemiec}, J., {Pohl}, M., {Stroman}, T., \& {Nishikawa}, K.-I. 2008, \apj,
  684, 1174

\bibitem[{{Ohira}(2013)}]{2013PhRvL.111x5002O}
{Ohira}, Y. 2013, \prl, 111, 245002

\bibitem[{{Ohira}(2016)}]{2016ApJ...827...36O}
---. 2016, \apj, 827, 36

\bibitem[{{Parizot} {et~al.}(2006){Parizot}, {Marcowith}, {Ballet}, \&
  {Gallant}}]{parizot_2006:snr_bampli}
{Parizot}, E., {Marcowith}, A., {Ballet}, J., \& {Gallant}, Y.~A. 2006, \aap,
  453, 387

\bibitem[{{Pohl} {et~al.}(2020){Pohl}, {Hoshino}, \&
  {Niemiec}}]{2020PrPNP.11103751P}
{Pohl}, M., {Hoshino}, M., \& {Niemiec}, J. 2020, Progress in Particle and
  Nuclear Physics, 111, 103751

\bibitem[{{Ressler} {et~al.}(2014){Ressler}, {Katsuda}, {Reynolds}, {Long},
  {Petre}, {Williams}, \& {Winkler}}]{Ressler2014}
{Ressler}, S.~M., {Katsuda}, S., {Reynolds}, S.~P., {Long}, K.~S., {Petre}, R.,
  {Williams}, B.~J., \& {Winkler}, P.~F. 2014, \apj, 790, 85

\bibitem[{{Reville} {et~al.}(2006){Reville}, {Kirk}, \&
  {Duffy}}]{2006PPCF...48.1741R}
{Reville}, B., {Kirk}, J.~G., \& {Duffy}, P. 2006, Plasma Physics and
  Controlled Fusion, 48, 1741

\bibitem[{{Sato} \& {Hughes}(2017)}]{2017ApJ...840..112S}
{Sato}, T., \& {Hughes}, J.~P. 2017, \apj, 840, 112

\bibitem[{{Scherer} {et~al.}(2020){Scherer}, {Baalmann}, {Fichtner},
  {Kleimann}, {Bomans}, {Weis}, {Ferreira}, \& {Herbst}}]{2020MNRAS.493.4172S}
{Scherer}, K., {Baalmann}, L.~R., {Fichtner}, H., {Kleimann}, J., {Bomans},
  D.~J., {Weis}, K., {Ferreira}, S.~E.~S., \& {Herbst}, K. 2020, \mnras, 493,
  4172

\bibitem[{{van Marle}(2020)}]{Vanmarle:2020}
{van Marle}, A.~J. 2020, \mnras, 496, 3198

\bibitem[{{van Marle} {et~al.}(2018){van Marle}, {Casse}, \&
  {Marcowith}}]{vanMarle:2018}
{van Marle}, A.~J., {Casse}, F., \& {Marcowith}, A. 2018, \mnras, 473, 3394

\bibitem[{{van Marle} {et~al.}(2019{\natexlab{a}}){van Marle}, {Casse}, \&
  {Marcowith}}]{vanMarle:2019}
---. 2019{\natexlab{a}}, \mnras, 490, 1156

\bibitem[{{van Marle} {et~al.}(2019{\natexlab{b}}){van Marle}, {Ryu}, {Kang},
  \& {Ha}}]{vanMarle2019b}
{van Marle}, A.~J., {Ryu}, D., {Kang}, H., \& {Ha}, J.-H. 2019{\natexlab{b}},
  Plasma and Fusion Research, 14, 4406119

\bibitem[{{Vay}(2008)}]{Vay2008}
{Vay}, J.~L. 2008, Physics of Plasmas, 15, 056701

\bibitem[{{Waxman} \& {Katz}(2017)}]{2017hsn..book..967W}
{Waxman}, E., \& {Katz}, B. 2017, {Shock Breakout Theory} (Springer
  International Publishing AG), 967

\bibitem[{{Wieland} {et~al.}(2016){Wieland}, {Pohl}, {Niemiec}, {Rafighi}, \&
  {Nishikawa}}]{2016ApJ...820...62W}
{Wieland}, V., {Pohl}, M., {Niemiec}, J., {Rafighi}, I., \& {Nishikawa}, K.-I.
  2016, \apj, 820, 62

\bibitem[{{Williams} {et~al.}(2013){Williams}, {Borkowski}, {Ghavamian},
  {Hewitt}, {Mao}, {Petre}, {Reynolds}, \& {Blondin}}]{Williams2013}
{Williams}, B.~J., {Borkowski}, K.~J., {Ghavamian}, P., {Hewitt}, J.~W., {Mao},
  S.~A., {Petre}, R., {Reynolds}, S.~P., \& {Blondin}, J.~M. 2013, \apj, 770,
  129

\bibitem[{{Xu} {et~al.}(2020){Xu}, {Spitkovsky}, \& {Caprioli}}]{Xu2020}
{Xu}, R., {Spitkovsky}, A., \& {Caprioli}, D. 2020, \apjl, 897, L41

\end{thebibliography}

\begin{appendix}
\section{A note on timescales}
One of the problems with comparing these simulations is that it is difficult to determine what constitutes 'the same moment in time'. There are four different timescales that need to be considered. 1) The dynamic timescale of the shock, which scales with $1/V_{\rm sh}$. 2) The response time of the upstream thermal plasma, which scales with $1/c_{\rm s}$. 3) The response time of the upstream magnetic field, which scales with $1/V_{\rm A}$. 4) The kinetic timescale of the non-thermal plasma, which sales with $1/v_{\rm p}$,  $v_{\rm p}$ being the representative particle velocity. The latter, in particular, is problematic, because the average particle velocity is time and space dependent. Near the shock, the recently injected particles dominate, which have a velocity of $v_{\rm inj}$. However, in the upstream medium, the only particles are those that have been accelerated by the SDA process. As a result, their velocity is typically about $3\,v_{\rm inj}$. 
The situation becomes even more complex when comparing simulations with different plasma-$\beta_{\rm p}$, which means that the timescales for gas and magnetic field in the upstream medium are no longer the same. 
For the sake of simplicity, we define the timescale as \ab{$\Omega_{\rm inj}^{-1}\,=\,r_{\rm l}/v{\rm inj}$} with $r_l$ the gyro radius defined by $v_{\rm inj}$ and $B_0$. However, we should keep in mind that, depending on the nature of the simulations, this does not guarantee synchronicity for all components of the plasma.

\section{The influence of the box-size}
\begin{figure}[!t]
\centering
\includegraphics[width=0.99\columnwidth]{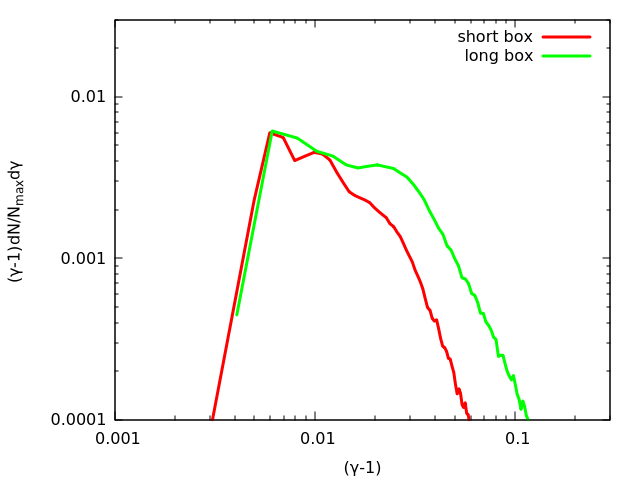}
\caption{Influence of box-length on the resulting SED.}
\label{fig:boxsize}
\end{figure}

As particles gain energy, they will travel further into the upstream medium before being reflected. Therefore, the size of the simulation box along the x-axis influenced the SED that is obtained. This was demonstrated in \cite{Vanmarle:2020}, which showed a much extended spectrum for a longer simulation box. However, in that case, even the small box simulation still showed that DSA took place. 
In our simulations, we are at the limit where DSA can take place at all and as a result, the box-size can also become a critical issue. To demonstrate this, we show the results of two simulations: PICMHD04, and a simulation that is identical, but in a box that has only half the length along the x-axis. 
As shown in Fig.~\ref{fig:boxsize}, the longer box-length clearly produces a more extended high-energy tail. Box-size limitations can be an issue if one want to investigate the raise of the maximum particle momentum as function of time.

\end{appendix}

\end{document}